\documentclass{an_art}
\usepackage{psfig}
\usepackage{graphicx}

\begin{document}

\begin{titlepage}
\setcounter{page}{000}
\headnote{Astron.~Nachr.~000 (0000) 0, 000--000}
\makeheadline

\title {The proper motion of the neutron star RXJ1856.5-3754
as measured by optical and X-ray imaging\thanks{Based
on observations obtained at the European Southern Observatory
on La Silla and Cerro Paranal in programs 59.D-0580, 63-H.0025,
and 65.H-0102, as well as additional archived VLT observations,
and with ROSAT} }

\author{{\sc R.~Neuh\"auser},
Garching, Germany \\
\medskip
{\small MPI f\"ur extraterrestrische Physik}
}

\date{Received ; accepted } 
\maketitle

\summary
The proper motion of the faint (V=25.6 mag) neutron star RXJ1856.5-3754
as measured consistently with both the FORS1 optical imager on the
ESO Very Large Telescope (VLT) as well as with the ROSAT High Resolution
Imager is presented.
With two VLT images obtained with a 0.95 year epoch difference
in Apr 1999 and Apr 2000, a proper motion of
$\mu _{\alpha} = 0.326 \pm 0.064 ^{\prime \prime}$ per year to the east
and no detectable motion in declination is found, consistent with
simultaneous findings by Walter (2001) using three HST observations.
Also, by comparing the positional differences between the strong X-ray
source RXJ1856.5-3754 and other X-ray sources detected in the same
ROSAT High Resolution Imager field (observed in Oct 1994 and Oct 1997),
a proper motion of $0.34 \pm 0.12 ^{\prime \prime}$ to the east is detected.
This may be the first star, where a proper motion
is clearly detected in X-ray images.
Additional VLT images taken in half-year intervalls give an upper limit
to the parallax of $\sim 51$ mas.
Furthermore, in archived VLT observations, the first B-band
detection of this object is found with B$=25.14 \pm 0.41$ mag.
END

\keyw
stars: neutron 
END
\end{titlepage}

\section{Introduction: RXJ1856.5-3754}

The X-ray source RXJ1856.5-3754 (or just RXJ1856)
was detected with the ROSAT All-Sky Survey and with ROSAT pointed observations.
Based on its strong and soft X-ray emission and its large X-ray to optical
flux ratio, Walter, Wolk, \& Neuh\"auser (1996) argued that RXJ1856 may be an
isolated neutron star (iNS).
Indeed, both Walter \& Matthews (1997) and Neuh\"auser et al. (1998)
found a faint ($V \simeq 25.6$ mag) blue star in the revised ROSAT HRI
error circle (Neuh\"auser et al. 1997).

It was expected to find thousands of such iNS with ROSAT (Colpi et al. 1993,
Madau \& Blaes 1994), but so far only a few were presented
(Haberl et al. 1997, 1998, 1999, Schwope et al. 1999, Motch et al. 1999).
This is probably due to the fact that (i) the velocity distribution assumed
in those theoretical estimations was found to be unrealistic in the meantime
(there are fewer slowly moving and hence less X-ray bright iNS),
and (ii) the optical counterparts of most ROSAT detected iNS are too faint
(scaling from RXJ1856, most will be fainter than 30 mag in V) for
optical detection and identification
(see Neuh\"auser \& Tr\"umper 1999 and Treves et al. 2000).

\section{Proper motion from ground-based optical imaging}

Neuh\"auser et al. (1998) presented the first ground-based detection,
where the 3.5m New Technology Telescope (NTT) of the European Southern
Observatory (ESO) in La Silla, Chile, was used with the ESO Multi-Mode Instrument
(EMMI). The pixel scale was $0.27 ^{\prime \prime}$ per pixel,
the total integration time was 150 minutes.
The NTT image will be used as 1st epoch image in the proper motion analysis.

Then, using the ESO 8.2m Antu, Unit Telescope No. 1 (UT1) of the
Very Large Telescope (VLT) on Cerro Paranal, Chile, RXJ1856
was observed three more times in the V-band, always using the
FOcal Reducer/low dispersion Spectrograph No. 1 (FORS1)
with its high resolution collimator ($0.1 ^{\prime \prime}$/pixel).
All VLT observations consist of seven co-added 15 min exposures.
See Table 1 for the observation log.

The VLT image taken in Apr 2000 is shown in figure 1.
The iNS RXJ1856 is clearly detected. The contours show the positions
of the same objects in the Aug 1997 NTT image.
One can clearly see that RXJ1856 moves to the east,
while the other objects do not move.

\begin{figure}
\resizebox{115mm}{!}
{\includegraphics[bbllx=70pt,bblly=303pt,bburx=522pt,bbury=755pt]{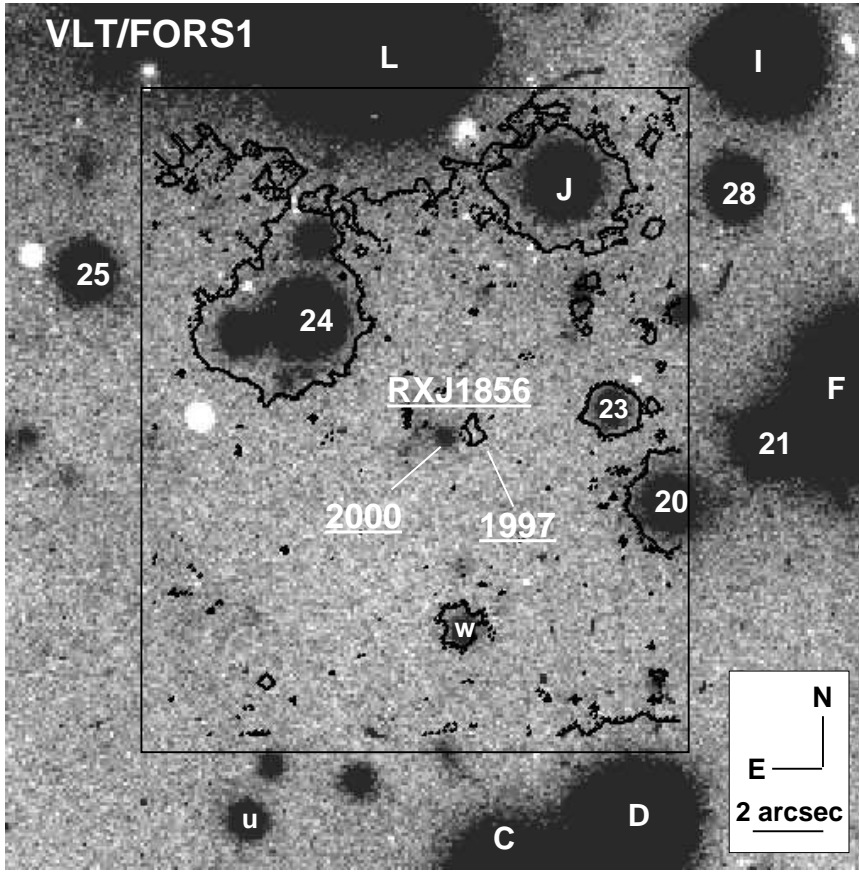}}
\hfill
\parbox[b]{55mm}{
\caption{Motion of RXJ1856 from 1997 to 2000.
The VLT image of RXJ1856 and surrounding area taken in Apr 2000 is shown.
Objects are labbeled as in Neuh\"auser et al. (1997).
Within the rectangular box, the contours of the Aug 1997 NTT image
are displayed (black lines). The faint iNS RXJ1856,
located in the center of the image, is clearly moving to the east.}
}
\end{figure}

The offsets between RXJ1856 and the stars I, J, 17, 19, 21, 23, 24, 26, 28,
and u as labelled in Neuh\"auser et al. (1997) were measured consistently
in all V-band images listed in table 1, to quantify the motion of RXJ1856.
The precision in the offset between the faint iNS RXJ1856
and any other object, for both $\alpha$ and $\delta$,
is $\sim 0.13 ^{\prime \prime}$ in the NTT image
and $\sim 0.047 ^{\prime \prime}$ in the VLT images.

By comparing the offsets in the NTT and the last VLT image,
a motion of $\Delta \alpha = 0.95 \pm 0.23 ^{\prime \prime}$
to the east and $\Delta \delta = 0.18 \pm 0.20 ^{\prime \prime}$
to the south during the 2.66 years epoch difference is found,
which should be a combination of proper and parallactic motion,
because the epoch difference is not an integer number of years.

\begin{table}
\begin{tabular}{cccccc}
\multicolumn{6}{c}{\bf Table 1. Observation log} \\ \hline
Date & {\scriptsize FWHM} & Tel. & Instr. & Expo. & \hspace{-.5cm} Band \\ \hline
10 Aug 1997  & $1.4 ^{\prime \prime}$ & NTT & EMMI & 150 min. & V \\
25 Apr 1999  & $0.6 ^{\prime \prime}$ & VLT & FORS1 & 105 min. & V \\
16 Jun 1999  & $0.9 ^{\prime \prime}$ & VLT & FORS1 & 20 min. & B \\
29 Aug 1999  & $0.7^{\prime \prime}$ & VLT & FORS1 & 105 min. & V \\
5,7 Apr 2000 & $0.5 ^{\prime \prime}$ & VLT & FORS1 & 105 min. & V \\ \hline
\end{tabular}
\end{table}

Because RXJ1856 is seen to move with the same proper motion
and in the same direction from image to image
(first with the NTT, then several times with the VLT),
a highly significant proper motion detection is obtained.
The B-band detection from Jul 1999 is not used in the proper motion
analysis, because RXJ1856 is only marginaly detected ($3 \sigma$).
Note the two faint (probably red) slightly extended objects
$\sim 1^{\prime \prime}$ east of RXJ1856 (figure 1), which were
also marginaly detected in the 1997 NTT V- and R-band images;
one of them is {\em star 115} in Walter (2001).
RXJ1856 is moving towards them, which will hamper optical
observations in the years 2003 to 2005.

The best and most precise proper motion estimate can be obtained
by comparing the offsets in the two VLT images taken
in Apr 1999 and Apr 2000 (epoch difference 0.95 years):
$\mu _{\alpha} = 0.326 \pm 0.064 ^{\prime \prime}$
per year to the east and
$\mu _{\delta} = 0.016 \pm 0.066 ^{\prime \prime}$ per year
to the north, i.e. basically to the east.

The detection of such a large proper motion is the final proof
for the correct identification of the strong X-ray source with
the faint optical counterpart, and hence the final proof
for the classification of RXJ1856 as iNS.

By observing RXJ1856 not only twice with a full-year epoch difference
for obtaining the proper motion, but also once in between, namely after
half a year, one can also try to detect parallactic motion.
However, given the precision in centroiding
the faint iNS in the VLT images ($\pm 40$ mas) and in the proper
motion, which has to be subtracted, no significant parallactic motion
can be detected. The upper limit is $\sim 51$ mas.

More or less simultaneously, Walter (2001) observed RXJ1856
three times in half-year intervalls with the HST,
also to measure proper motion and parallax. He found a proper motion
of $\mu _{\alpha} = 0.3267 \pm 0.0008 ^{\prime \prime}$ to the east
and $\mu _{\delta} = 0.0591 \pm 0.0007 ^{\prime \prime}$ to the south,
perfectly consistent with the ground-based result.
The fact that the ground-based VLT result is consistent with Walter's HST
result also shows that Walter (2001) dealt correctly with the remnant
plate distortions and pixel scale variations on the HST detector.
Walter (2001) also detected the parallax of RXJ1856 being
$16.5 \pm 2.3$ mas corresponding to a distance of only $\sim 61$ pc,
so that RXJ1856 is the nearest known neutron star.

\section{Proper motion from X-ray imaging with ROSAT}

RXJ1856 was observed by the ROSAT High Resolution Imager (HRI)
in Oct 1994 and Oct 1997 with 3.012 years epoch difference (PI F.M. Walter),
so that it can be used to measure the proper motion.
The first HRI observation was discussed before by Neuh\"auser et al. (1997),
where the exact HRI position and its error circle was determined.
Both data sets were reduced in the standard way using MIDAS/EXSAS
to obtain sources lists and source positions.
Note that two HRI pixels correspond to $0.9957 ^{\prime \prime}$
(David et al. 1996). For more details on ROSAT, see Tr\"umper (1982).

Several sources are detected in the field with RXJ1856 being the brightest.
The positional offsets between RXJ1856 and the other sources in the HRI
field were determined after individual boresight correction.
Compared to the other non-moving sources,
RXJ1856 moved by about one arc sec to the
east in 3.012 years, corresponding to a proper motion of
$\mu _{\alpha} = 0.34 \pm 0.12 ^{\prime \prime}$ per year to the east,
consistent with the optical measurements.

This may be the first case, where a proper motion of a stellar object was
measured by X-ray imaging (as noted before by Walter \& An 1998).
Even though the nominal positional accuracy of the ROSAT HRI is only
$3.8 ^{\prime \prime}$, a measurement of this much smaller motion
was possible, because RXJ1856 is such a strong X-ray source
and because there are several other X-ray sources around
RXJ1856 in the same HRI field.

\section{Deep imaging in the B-band}

RXJ1856 was observed four times five minutes in June 1999 in the
Bessel B-band with the FORS1 large field at the VLT-Antu,
and the data were taken from the public VLT archive.
After co-adding the images, the conversion from instrumental magnitude
to apparent B-band magnitude was estimated by using the stars C, D, F, I,
J, L, 19, 24, 25, 26, and 28, whose B-band magnitudes are given
in Neuh\"auser et al. (1997).
The error in the B magnitude of RXJ1856 comes from the error in the above
conversion ($\pm 0.32$ mag) and from the noise ($\pm 0.25$ mag)
in the marginal detection ($3 \sigma$) of the faint object RXJ1856.
We obtain B$=25.14 \pm 0.41$ mag. As seen in figure 2,
the F300 (HST U), B, F606 (HST V), and V-band fluxes do not show
any evidence for an optical emission or absorption feature,
but they are above the 57 eV blackbody fit to the ROSAT PSPC spectrum.
This may indicate that the RXJ1856 atmosphere is composed of
elements higher than H or He, e.g. Si or Fe 
(c.f. Pavlov et al. 1996, Pons et al. 2000).

\begin{figure}
\vbox{\psfig{figure=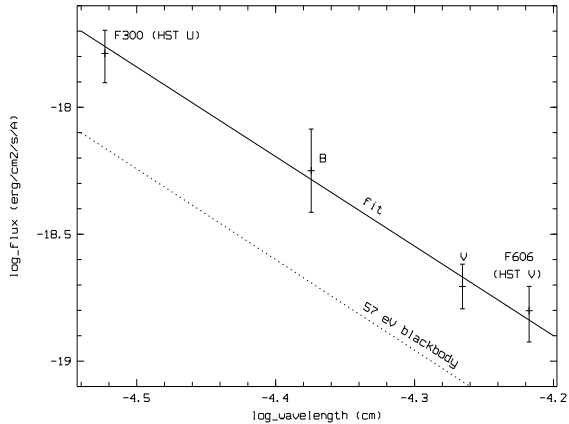,height=6cm,width=8cm,angle=270}}
\caption{Optical spectral energy distribution of RXJ1856.
The HST F300 and F606 fluxes (Walter \& Matthews 1997) are shown
together with the NTT V-band (Neuh\"auser et al. 1998)
and VLT B-band flux (this paper).
The dotted line shows the Rayleigh-Jeans extention of the
unreddened 57 eV blackbody fit to the ROSAT PSPC flux
(Walter, Wolk, \& Neuh\"auser 1996), which is below the
optical fluxes by a factor of 2 to 3.
The full line shows the best fit to the optical data,
probably indicating a somewhat higher-element composition.}
\end{figure}

\section{Discussion and outlook}

The proper motion of the iNS RXJ1856 as measured consistently with optical images
taken with NTT and VLT as well as with ROSAT X-ray images is found to be $\sim 1/3$
of an arc sec per year to the east.
At the distance of RXJ1856 ($\sim 61$ pc as measured by HST, Walter 2001)
this corresponds to a velocity of $\sim 88$ km/s.
This velocity is too large for RXJ1856 being an old iNS accreting
from the interstellar medium to achieve its strong X-ray emission:
According to Bondi-Hoyle accretion (scaled with $v ^{-3}$),
its velocity $v$ would need to be as small as $\sim 5$ km/s
for a normal density of the interstellar medium of $\sim 1~cm^{-3}$.
A much higher density is excluded from the basically unabsorbed
($A_{V} \simeq 0.07$ mag) ROSAT X-ray spectrum.
Hence, RXJ1856 is a relatively young iNS with a cooling,
but still hot surface producing strong X-ray emission.

In a recent ESO press 
release\footnote{see 
http://www.eso.org/outreach/press-rel/pr-2000/pr-19-00.html }
an H$\alpha$ emitting region was shown around RXJ1856, cone-shaped
with RXJ1856 near its edge, the major axis alinged with its 
proper motion. 
It was argued that the H$\alpha$ emission is an ionization front
originating from a dense nebula, 
from which RXJ1856 accretes to generate its strong 
X-ray emission. However, the shape of the emission region is
also prefectly consistent with a bow shock, expected
from a fast-moving nearby neutron star like RXJ1856
(see Walter et al., in preparation, for more details).

Tracing back the proper motion of RXJ1856, it moved through
(or in front of or behind) the Sco-Oph star forming region 
some 1 Myrs ago, from where the run-away
O star $\zeta$ Oph has been ejected. 
Such a chance alignment is very unlikely.
Hence, Walter (2001) argues that both stars formed a binary system
and were then ejected to opposite
directions by the supernova of the progenitor of RXJ1856.
A particular radial velocity is required,
so that RXJ1856 really moved through the Oph star forming
region, instead of passing behind or before. 
Then, both objects were at the same location
in the Upper Sco OB association $\sim 1.1$ Myrs ago
(Walter 2001), recently revised to $\sim 0.9$ Myrs 
(Walter 2000), which would be the age of RXJ1856.

As nearby iNS with strong X-ray emission, RXJ1856 is well suited for
X-ray spectroscopy with Chandra and XMM-Newton, in order to determine
the atmospheric composition (and hence the radius from the X-ray flux)
as well as the gravitational redshift and surface gravity,
both proportional to mass and radius. Altogether, these data will
constrain the nuclear equation-of-state (c.f.
Neuh\"auser et al. 2000).
The Chandra observation (part of MPE's garanteed time by P. Predehl
and R.N.) was executed and the XMM-Newton observation (part of
MPE's garanteed time by B. Aschenbach and R.N.) will be executed soon.

\acknowledgements
I would like to thank Fred M. Walter for useful discussion,
providing me with the second ROSAT HRI data set, 
and for sending me his recent paper
on RXJ1856 before publication. Also, I am grateful to the NTT team on La Silla
for their help, as well as to the ESO staff on Cerro Paranal
for their service mode observation at good seeing conditions,
and I would also like to thank the ESO User Support Group for assistance.
I would also like to acknowledge H.-C. Thomas 
and F. Comer\'on for their contributions.
The ROSAT project is supported by the German 
Government (BMBF/DLR) and the Max-Planck-Society.
I would like to acknowledge financial support from the Bundesministerium 
f\"ur Bildung und
Forschung through the Deutsche Zentrum f\"ur Luft- 
und Raumfahrt e.V. (DLR) under
grant number 50 OR 0003.

\refer
\aba
\rf{Colpi M., Campana S., Treves A., 1993, A\&A 278, 161}
\rf{David L.P., Harnden F.R., Kearns K.E., Zombeck M.V., 1996, The ROSAT
High Resolution Imager calibration report. SAO technical report}
\rf{Haberl F., Motch C., Buckley D.A.H., Zickgraf F.-J.,
Pietsch W., 1997, A\&A 326, 662}
\rf{Haberl F., Motch C., Pietsch W., 1998, AN 319, 97}
\rf{Haberl F., Pietsch W., Motch C., 1999, A\&A 351, L53}
\rf{Madau P. \& Blaes O., 1994, ApJ 423, 748}
\rf{Motch C., Haberl F., Zickgraf F.-J., Hasinger G.,
Schwope A.D., 1999, A\&A 351, 177}
\rf{Neuh\"auser R. \& Tr\"umper J., 1999, A\&A 143, 151}
\rf{Neuh\"auser R., Thomas H.-C., Danner R., Peschke S., Walter F.M.,
1997, A\&A 318, L43}
\rf{Neuh\"auser R., Thomas H.-C., Walter F.M., 1998, The Messenger 92, 27}
\rf{Neuh\"auser R., Burwitz V., Zavlin V.E., Predehl P., 
Tr\"umper J., Walter F.M., 2000, BAAS abstract HEAD 32.3605}
\rf{Pfeffermann E., et al., 1988, The focal plane instrumentation of the
ROSAT telescope. In: Proc. SPIE 733, 519}
\rf{Pavlov G.G., Zavlin V.E., Tr\"umper J., Neuh\"auser R., 
1996, ApJ 472, L33}
\rf{Pons J.A., Walter F.M., Lattimer J.M., Prakash M.,
2000, BAAS abstract HEAD 32.3314}
\rf{Schwope A.D., Hasinger G., Schwarz R. Haberl F., Schmidt M.,
1999, A\&A 341, L51}
\rf{Treves A., Turolla R., Zane S., Colpi M., 2000, PASP 112, 297}
\rf{Tr\"umper J., 1982, Adv. Space Res. 2 (no. 4), 241}
\rf{Walter F.M., 2000, BAAS abstract HEAD 32.3313}
\rf{Walter F.M., 2001, ApJ, in press (10 Jan 2001), astro-ph/0009031}
\rf{Walter F.M. \& An P., 1998, BAAS 192, 50.04}
\rf{Walter F.M. \& Matthews L.D., 1997, Nature 389, 358}
\rf{Walter F.M., Wolk S.J., Neuh\"auser R., 1996, Nature 379, 233}
\abe

%
\address 
\rf{Ralph Neuh\"auser, MPI f\"ur Extraterrestrische Physik, D-85740 Garching, Germany, rne@mpe.mpg.de}

\end{document}